\documentclass[conference]{IEEEtran}
\IEEEoverridecommandlockouts

\usepackage{cite}
\usepackage{amsmath,amssymb,bm}
\usepackage{algorithm}
\usepackage{algorithmic}
\usepackage{graphicx}
\usepackage{booktabs}
\usepackage{array}
\usepackage{xcolor}
\usepackage{placeins}
\usepackage{balance}
\usepackage{url}
\usepackage[colorlinks]{hyperref}
\hypersetup{
  citecolor=blue,
  linkcolor=red,
  urlcolor=blue
}
\newcommand{\herm}{\mathrm{H}}
\newcommand{\trans}{\mathrm{T}}
\newcommand{\C}{\mathbb{C}}
\newcommand{\R}{\mathbb{R}}
\newcommand{\junit}{\mathrm{j}}
\newcommand{\norm}[1]{\left\lVert#1\right\rVert}

\title{Off-Grid Position Optimization under Mutual Coupling in Fluid Antenna Arrays}

\author{Jingyuan Xu, Jian Dang, Zaichen Zhang
  \thanks{Jingyuan Xu is with the National Mobile Communications Research Laboratory, Frontiers Science Center for Mobile Information Communication and Security, Southeast University, Nanjing 210096, China (e-mail: jingyuanxu@seu.edu.cn).}

  \thanks{Zaichen Zhang is with the National Mobile Communications Research Laboratory, Frontiers Science Center for Mobile Information Communication and Security, Southeast University, Nanjing 210096, China. Zaichen Zhang is also with Purple Mountain Laboratories, Nanjing 211111, China. (e-mail: zczhang@seu.edu.cn).}

	\thanks{Jian Dang is with the National Mobile Communications Research Laboratory, Frontiers Science Center for Mobile Information Communication and Security, Southeast University, Nanjing 210096, China, also with the Key Laboratory of Intelligent Support Technology for Complex Environments, Ministry of Education, Nanjing University of Information Science and Technology, Nanjing 210044, China, and also with Purple Mountain Laboratories, Nanjing 211111, China. (e-mail: dangjian@seu.edu.cn).}

}

\begin{document}
\raggedbottom
\bstctlcite{IEEEexample:BSTcontrol}
\maketitle
\pagestyle{plain}
\thispagestyle{plain}

\begin{abstract}
Fluid antenna arrays exploit continuous antenna repositioning within a finite aperture to provide geometry diversity beyond grid-constrained port selection. Every displacement, however, changes both the radiation response and the multiport mutual-impedance network, coupling geometry optimization with the source-voltage constraint. This paper develops an electromagnetic-aware (EM-aware) beamforming framework for planar fluid antenna arrays. Phase retrieval converts an amplitude-only shaped-beam specification into an aperture-compatible complex target, and an EM-aware orthogonal matching pursuit (OMP) method selects grid-constrained initial antenna positions. Continuous refinement then alternates exact voltage-constrained current optimization with movement-constrained projected adaptive moment estimation (Adam) updates of all physical antenna positions. Across independently perturbed symmetric dual-beam targets, the proposed method consistently improves the average mainlobe signal-to-noise ratio (SNR) and reduces the peak sidelobe level (PSLL) over a uniform array and discrete port selection.
\end{abstract}


\section{Introduction}\label{sec:introduction}

\subsection{Background and Related Work}

Beamforming is the mechanism by which an array of simple radiating
elements acquires spatial selectivity.  Through the coherent superposition
of waves radiated by multiple elements, the array concentrates energy
toward intended directions while suppressing it elsewhere, so that the
beam pattern directly shapes the interference level, the coverage, and the
energy efficiency of the link.  In conventional designs, however, the
element geometry is fixed during deployment, and only the feed
coefficients are optimized~\cite{nb1,nb2}.  A demanding specification,
such as uniform coverage over a macro
sector~\cite{nb2}, is then met by scaling the array, adding feeds, or
accepting a compromise among conflicting masks, because the fixed geometry
bounds the achievable patterns no matter how the weights are chosen.
Allowing the radiation positions themselves to move would therefore enlarge
the pattern design space substantially, and this flexibility is the central
idea underlying fluid antenna architectures.

Fluid antenna systems (FASs)\cite{wong2020limits,wong2021fas,zhang2026fbl} provide exactly this flexibility at the
antenna level, where the active radiating element is no longer anchored at
a fixed location but can be relocated or switched among candidate
positions inside a finite
aperture.  The term
``fluid'' describes this softwarized adaptability, not liquid
antenna materials, and the effective radiation position is therefore
electronically controllable.  Early studies focused on the spatial
fluctuation of the channel across the aperture, the so-called random
{\em fading gain}~\cite{wt2,zzt1,zzt_wcl4,wt5,zzt_wcl1,zzt_wcl2,hj3,zzt_wcl3}.  Because the
channel envelope takes a different value at every candidate position,
selecting the most favorable port yields a selection-diversity gain,
improving the point-to-point link without additional radio-frequency
chains. Recent works have been investigating the channel reconstruction limits with feasible practical designs~\cite{CSI1,CSI2,CSI3,CSI4,CSI5}.

A complementary and more fundamental dimension of FAS
has recently emerged, the geometric diversity of a fluid
antenna array (FAA), which activates multiple ports
simultaneously~\cite{zhang2026finite,zhang2026planar}.  In contrast to
fading-based benefits, this diversity is easy to exploit,
because it comes from a deliberately reconfigurable geometric structure.  The
reconfigurable geometry
directly determines the effective aperture, the spatial sampling structure,
and the spatial frequency content of the array, and hence its radiation
pattern, so the achievable performance is qualitatively different from
that of fixed apertures.  For
both linear and
planar FAA topologies~\cite{zhang2026finite,zhang2026planar}, this geometric
flexibility enables fine-grained radiation
pattern control, including beamforming gain enhancement, directivity
shaping, and, most importantly for interference-limited systems, sidelobe
suppression.  When candidate positions become densely arranged, mutual
coupling must be included in the model, which has led to
electromagnetic-aware (EM-aware) FAA designs~\cite{zhang2026emaware}.
Designing the port activation of an FAA is thus a geometry-aware
array synthesis problem, not just a channel selection problem.
Driven by this perspective, recent designs include flexible
arbitrary-direction beam synthesis~\cite{11555734}, near-field
beamforming and port selection~\cite{chen2026nearfield}, peak-sidelobe
suppression~\cite{liang2026peaksidelobe}.

Selecting a small set of active ports from a dense candidate grid is a
sparse-approximation problem.  Compressive-sensing-based schemes
approximate the desired field with a few columns of the array-response
dictionary, greedily selecting the most correlated column and refitting
the selected coefficients.  Such schemes are attractive whenever the
number of radio-frequency feeds is far smaller than the number of
candidate positions.  Grid selection alone, however, cannot exploit
positions between candidates, and off-grid sparse-array results indicate
that continuous location refinement removes the basis mismatch caused by
grid quantization~\cite{yang2022sparse}.  When the candidate positions
become densely arranged, electromagnetic mutual coupling changes the
relationship among terminal voltages, antenna currents, and radiated
fields~\cite{gupta1983mutual,balanis2016}, so that the array must be
modeled as a coupled multiport network and the support and beamforming
weights must be designed under this coupled response~\cite{zhang2026emaware}.
In the continuous design adopted here, the grid serves only as a numerical
initializer, and the coupled network is rebuilt as the selected antennas
move.

\subsection{Challenges and Motivations}

Three challenges motivate a dedicated continuous-position design.
\begin{itemize}
\item \textbf{Support-dependent atoms:} In standard OMP, the dictionary atoms are fixed and independent of the selected support~\cite{tropp2007omp}, whereas EM-aware modeling embeds the coupling network into the beamformer~\cite{zhang2026emaware}.  Because adding a candidate changes the impedance network of the selected physical antennas, its effective field atom depends on the current support and cannot be obtained from a fixed full-grid coupled dictionary.
\item \textbf{Current--position coupling:} In coupling-aware synthesis, the array geometry is fixed and only the currents are optimized~\cite{echeveste2016,zhang2011robust}, while mutual coupling changes the map from source voltages to port currents~\cite{gupta1983mutual}.  When the geometry is optimized as well, moving one antenna changes the steering response, mutual impedances, optimal port currents, and the source-voltage-constrained feasible set.  Holding the currents or coupling network fixed therefore gives an inconsistent geometry gradient.
\item \textbf{Constrained motion:} Grid-based designs keep ports on a discrete lattice and avoid continuous feasibility checks~\cite{11555734}.  Off the lattice, the aperture, minimum pairwise spacing, and per-antenna travel disks define a nonconvex feasible set.  Position updates must preserve these physical constraints while moving all antennas and repeatedly rebuilding the coupled network.
\end{itemize}

\subsection{Contributions}

The contributions are summarized as follows.
\begin{itemize}
\item \textbf{EM-aware initialization:} Phase retrieval based on the Gerchberg--Saxton method~\cite{gerchberg1972,11555734} constructs an aperture-compatible complex target.  A batched EM-aware OMP then evaluates support-dependent candidate innovations and produces a grid-constrained physical initialization.
\item \textbf{Exact current and reduced gradient:} For every fixed geometry, the voltage-constrained current subproblem is solved exactly.  An envelope-theorem reduced gradient then incorporates both the radiation-response derivative and the mutual-impedance sensitivity of the source-voltage constraint.
\item \textbf{Projected Adam:} Projected adaptive moment estimation (Adam) jointly moves all physical antennas while cyclically enforcing aperture, travel, and minimum-spacing constraints.  Experiments on independent targets and movement budgets quantify its gains over EM-aware OMP and a uniform array.
\end{itemize}

\emph{Notation:} Bold lowercase and uppercase symbols denote vectors and matrices, and calligraphic symbols denote sets.  For a matrix $\mathbf A$, $[\mathbf A]_{m,n}$ is its $(m,n)$th entry, and $[\mathbf A]_{\mathcal I,\mathcal J}$ selects rows $\mathcal I$ and columns $\mathcal J$; a colon retains all indices and a scalar index is a singleton.  The cardinality of $\mathcal I$ is $|\mathcal I|$.  The superscripts $(\cdot)^{\trans}$ and $(\cdot)^{\herm}$ denote transpose and Hermitian transpose.  The sets $\R$ and $\C$ denote real and complex numbers, $\junit=\sqrt{-1}$, $\Re\{\cdot\}$ takes the real part, $\odot$ is elementwise multiplication, $\norm{\cdot}_2$ is the Euclidean norm, and $\mathbf I$ is an identity matrix of context-dependent size.

\section{System Model}\label{sec:model}

\subsection{Continuous Position Optimization}

\begin{figure}[t!]
\centering
\includegraphics[width=\columnwidth]{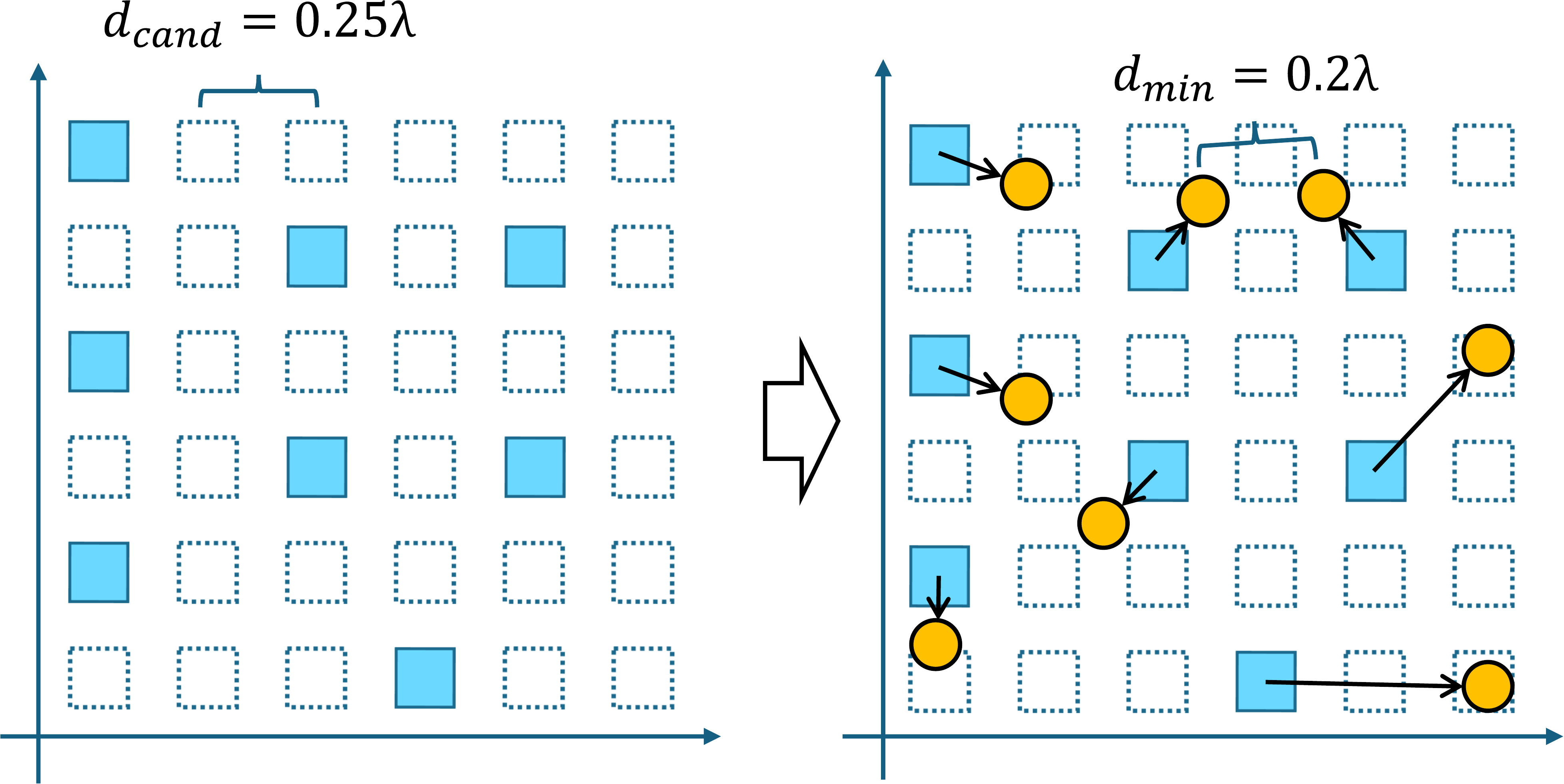}
\caption{Geometry of the planar FAA.  The candidate lattice has spacing $d_{\rm cand}$ (left), while the physical ports occupy continuous positions with minimum pairwise spacing $d_{\min}$ (right).  Arrows indicate displacement (not to scale).}
\label{fig:continuous_structure}
\vspace{-2mm}
\end{figure}

Consider a planar FAA comprising $N_a$ physical, parallel half-wave dipoles in the $x$--$y$ plane.  Their positions are
\begin{equation}
 \mathbf R=[\mathbf r_1,\ldots,\mathbf r_{N_a}]^{\trans},~
 \mathbf r_n=[x_n,y_n]^{\trans}\in\mathcal A,
\end{equation}
where $\mathbf R\in\R^{N_a\times2}$ is the array-geometry matrix, $\mathbf r_n$ is antenna $n$'s position, and $x_n$ and $y_n$ are its Cartesian coordinates.  The feasible aperture $\mathcal A=[0,L]^2$ is a square of side length $L$.  A uniform $N_p$-point candidate-port lattice $\mathcal P=\{\mathbf p_q\}_{q=1}^{N_p}\subset\mathcal A$ has nearest-neighbor spacing $d_{\rm cand}$, where $\mathbf p_q\in\R^2$ denotes candidate coordinate $q$.  The physical geometry satisfies the minimum pairwise spacing constraint $\norm{\mathbf r_n-\mathbf r_m}_2\ge d_{\min}$ for all $n\ne m$.  Fig.~\ref{fig:continuous_structure} illustrates these geometric quantities.

Following the RMS-phasor convention in~\cite{zhang2026emaware}, let $i_n$ denote the feed-current phasor at the center-fed terminal of antenna $n$, and collect these currents as $\mathbf i=[i_1,\ldots,i_{N_a}]^{\trans}\in\C^{N_a}$.  Write $i_n=|i_n|\exp(\junit\psi_n)$, where $|i_n|$ and $\psi_n=\angle i_n$ are its amplitude and phase.  The square direction-cosine design grid has $N_{\rm d}$ samples per axis, of which $M$ lie in the visible disk.  For visible far-field sample $\ell$ at $(\theta_\ell,\phi_\ell)$, define the spatial frequencies $u_\ell=\sin\theta_\ell\cos\phi_\ell/\lambda$ and $v_\ell=\sin\theta_\ell\sin\phi_\ell/\lambda$.  Let $f_\ell$ denote the known direction-dependent factor containing the half-wave-dipole pattern, polarization, and dimensional normalization.  The current-to-radiation response matrix $\mathbf B(\mathbf R)\in\C^{M\times N_a}$ is then given by
\begin{equation}
 \bigl[\mathbf B(\mathbf R)\bigr]_{\ell,n}
 =f_\ell\exp[-\junit2\pi(x_nu_\ell+y_nv_\ell)].
\label{eq:current_response}
\end{equation}
The radiated field at sample $\ell$ is the coherent sum of all port-current contributions:
\begin{equation}
\begin{aligned}
 y_\ell
 &=\sum_{n=1}^{N_a}[\mathbf B(\mathbf R)]_{\ell,n}i_n\\
 &=f_\ell\sum_{n=1}^{N_a}|i_n|
 \exp\!\left\{\junit\!\left[\psi_n-2\pi(x_nu_\ell+y_nv_\ell)\right]\right\}.
\end{aligned}
\label{eq:current_field}
\end{equation}
Thus, $\mathbf y=\mathbf B(\mathbf R)\mathbf i$ explicitly couples the amplitude and phase of every port current with its continuous position.

\subsection{Electromagnetic Network Mutual Coupling}

Let $\mathbf Z_{\rm em}(\mathbf R)\in\C^{N_a\times N_a}$ be the reciprocal electromagnetic impedance matrix.  Let $d_{m,n}=\norm{\mathbf r_m-\mathbf r_n}_2$ denote the separation between antennas $m$ and $n$, and let $z(d)$ be the scalar mutual-impedance law.  Its diagonal contains the half-wave self impedance and a fixed real passivity loading $\rho_z$, whereas its off-diagonal entry $[\mathbf Z_{\rm em}]_{m,n}=z(d_{m,n})$ is the induced-EMF mutual impedance between antennas $m$ and $n$~\cite{gupta1983mutual,balanis2016}.  The small $\rho_z\mathbf I$ loading removes numerical passivity violations of the analytical mutual-impedance approximation.  With ohmic loss resistance $r_{\rm loss}$ and source impedance $z_s$, define the loaded network matrix
\begin{equation}
 \mathbf C(\mathbf R)=\mathbf Z_{\rm em}(\mathbf R)
 +(r_{\rm loss}+z_s)\mathbf I.
\label{eq:C}
\end{equation}
Here $\mathbf I\in\R^{N_a\times N_a}$.  Let $\mathbf v_s\in\C^{N_a}$ denote the open-circuit source-voltage phasor vector.  The network maps antenna-port currents to source voltages as
\begin{equation}
 \mathbf v_s=\mathbf C(\mathbf R)\mathbf i.
\label{eq:vi}
\end{equation}
Thus, currents are the beam-fitting variables, while the source voltages follow from the network and their norm is a physical constraint.

Let $\mathbf g_{\mathrm{am}}\in\R_+^M$ denote the desired beam-amplitude vector, and let $\mathbf g\in\C^M$ denote the corresponding phase-aware desired beam.  For beam $b$, let $\mathcal B_b$ contain its design samples, set $n_b=|\mathcal B_b|$, and define the balancing factor $\eta_b=(\max_j n_j/\max\{n_b,1\})^\gamma$.  Set $\bar w_m=w_{\rm side}$ when $[\mathbf g_{\rm am}]_m\le0.05$ and $\bar w_m=1$ otherwise.  The final weight is $w_m=\max(\{\bar w_m\}\cup\{\eta_b:m\in\mathcal B_b\})$, and $\mathbf W=\operatorname{diag}(w_1,\ldots,w_M)\succeq0$.  Thus, $w_{\rm side}$ emphasizes sidelobes, while $\gamma$ balances beams having different sample counts.  Let $\rho_i>0$ be a small current-regularization coefficient.  The joint design is
\begin{equation}\label{eq:joint}
\begin{aligned}
 \min_{\mathbf R,\mathbf i}~&
 \norm{\mathbf W^{1/2}(\mathbf B(\mathbf R)\mathbf i-\mathbf g)}_2^2
 +\rho_i\norm{\mathbf i}_2^2,\\
 \mathrm{s.t.}~&\norm{\mathbf C(\mathbf R)\mathbf i}_2\le V_{\max},\\
 &\mathbf r_n\in\mathcal A,~\forall n,\\
 &\norm{\mathbf r_n-\mathbf r_m}_2\ge d_{\min},~\forall n\ne m,\\
 &\norm{\mathbf r_n-\mathbf r_n^{(0)}}_2\le D,~\forall n.
\end{aligned}
\end{equation}
Here $V_{\max}$ is the source-voltage budget, $d_{\min}$ is the minimum pairwise spacing, $D$ is the per-antenna travel budget, and $\mathbf r_n^{(0)}$ is antenna $n$'s reference position.  The objective fits the complex field while regularizing current magnitude.  Because $\mathbf C(\mathbf R)$ varies with geometry, the voltage constraint is also position dependent, making steering-only refinement inadequate.
\begin{algorithm}[t!]
	\caption{Off-Grid Position Optimization Framework}
	\label{alg:adam}
	\begin{algorithmic}[1]
		\REQUIRE $\mathbf g_{\mathrm{am}}$, $\mathcal P$, $N_a$, $\mathbf W$,
		$\rho_i$, and $V_{\max}$
		\REQUIRE $L$, $d_{\min}$, $D$, $\alpha$, $\beta_1$, $\beta_2$,
		$\epsilon$, $\Delta_{\rm step}$, and $T_{\max}$
		\STATE Recover the phase and construct $\mathbf g$ from $\mathbf g_{\mathrm{am}}$
		\STATE Run EM-aware OMP and obtain $\mathbf R^{(0)}$
		\STATE Set $\mathbf m_0=\mathbf 0$, $\mathbf s_0=\mathbf 0$
		\STATE Solve the inner problem in~\eqref{eq:fixed_current} at
		$\mathbf R^{(0)}$ for $\mathbf i_0$
		\STATE Set $F_0\leftarrow F(\mathbf R^{(0)})$
		\STATE $(\mathbf R_{\rm best},\mathbf i_{\rm best},F_{\rm best})\leftarrow
		(\mathbf R^{(0)},\mathbf i_0,F_0)$
		\FOR{$t=1,\ldots,T_{\max}$}
		\STATE Form $\nabla F(\mathbf R_{t-1})$ from~\eqref{eq:gradient}
		\STATE Update and bias-correct the moments using~\eqref{eq:adam_moments}
		\STATE Form~\eqref{eq:adam_position} and cap the step at
		$\Delta_{\rm step}$
		\STATE $\mathbf R_t\leftarrow
		\mathcal P_{\rm ap,tr,sp}(\widetilde{\mathbf R}_t;\mathbf R^{(0)})$
		\STATE Rebuild the network and solve~\eqref{eq:fixed_current} for
		$\mathbf i_t$
		\STATE Set $F_t\leftarrow F(\mathbf R_t)$
		\IF{$F_t<F_{\rm best}$}
		\STATE $(\mathbf R_{\rm best},\mathbf i_{\rm best},F_{\rm best})\leftarrow
		(\mathbf R_t,\mathbf i_t,F_t)$
		\ENDIF
		\ENDFOR
		\RETURN $(\mathbf R_{\rm best},\mathbf i_{\rm best})$
	\end{algorithmic}
\end{algorithm}
\section{Proposed Method}\label{sec:method}

Problem~\eqref{eq:joint} is nonconvex in the antenna positions, while its
current subproblem is convex for every fixed geometry.  We therefore split the
design into discrete initialization and continuous refinement.  Phase retrieval
and an EM-aware OMP
first produce grid-constrained initial positions.  The selected ports are then
treated as the only physical antennas, and continuous refinement alternates
between exact current optimization and an EM-aware geometry
update.

\subsection{Phase Retrieval and EM-Aware OMP Initialization}

A phase-retrieval algorithm based on the Gerchberg--Saxton
method~\cite{gerchberg1972,11555734} recovers $\bm\psi\in\R^M$ and
constructs the phase-aware desired beam
\begin{equation}
 \mathbf g=\mathbf g_{\mathrm{am}}\odot\exp(\junit\bm\psi),
 \qquad
 \widetilde{\mathbf g}=\mathbf W^{1/2}\mathbf g .
\label{eq:phase_target}
\end{equation}
On the candidate grid, let $\mathbf B_{\mathcal P}\in\C^{M\times N_p}$
collect the isolated current-to-radiation response columns, let
$\overline{\mathbf B}_{\mathcal P}=\mathbf W^{1/2}\mathbf B_{\mathcal P}$,
and let $\mathbf C_{\mathcal P}\in\C^{N_p\times N_p}$ collect the loaded self
and pairwise mutual impedances.  These are numerical lookup tables, not an
$N_p$-antenna physical structure.  For a selected support
$\mathcal S\subseteq\{1,\ldots,N_p\}$, define
$\overline{\mathbf B}_{\mathcal S}
=[\overline{\mathbf B}_{\mathcal P}]_{:,\mathcal S}$ and
$\mathbf C_{\mathcal S,\mathcal S}
=[\mathbf C_{\mathcal P}]_{\mathcal S,\mathcal S}$.  Only the latter is the
physical network associated with $\mathcal S$.  Its support-dependent dictionary
\begin{equation}
 \mathbf D_{\mathcal S}
 =\overline{\mathbf B}_{\mathcal S}\mathbf C_{\mathcal S,\mathcal S}^{-1}
\label{eq:support_dictionary}
\end{equation}
maps auxiliary source voltages to weighted fields.

Let
$\widehat{\mathbf v}_{\mathcal S}$ be the regularized least-squares
coefficient vector and
\(
\mathbf r_{\mathcal S}
=\widetilde{\mathbf g}-\mathbf D_{\mathcal S}
\widehat{\mathbf v}_{\mathcal S}
\)
the residual.  For an unselected candidate index
$c\in\{1,\ldots,N_p\}\setminus\mathcal S$, define
$\overline{\mathbf b}_c=[\overline{\mathbf B}_{\mathcal P}]_{:,c}$ and
$\mathbf c_{\mathcal S,c}=[\mathbf C_{\mathcal P}]_{\mathcal S,c}$.
Block inversion of the trial network gives an added effective column
proportional to the innovation direction
\begin{equation}
 \mathbf q_c=\overline{\mathbf b}_c-
 \overline{\mathbf B}_{\mathcal S}
 \mathbf C_{\mathcal S,\mathcal S}^{-1}\mathbf c_{\mathcal S,c}.
\label{eq:schuratom}
\end{equation}
The proportionality factor is the inverse scalar Schur complement, which
cancels from the normalized correlation.  The next location therefore
maximizes
\(
|\mathbf q_c^{\herm}\mathbf r_{\mathcal S}|/\norm{\mathbf q_c}_2
\).
All candidates are evaluated in a batch using the Gram matrix of the
response columns and their cross-correlations with the residual.  After
$N_a$ selections, the chosen coordinates form $\mathbf R^{(0)}$.  All
unselected candidate points are discarded and do not participate in the
continuous-stage electromagnetic network.

\subsection{Exact Voltage-Constrained Current Optimization}

For fixed $\mathbf R$, define
\(
\mathbf A(\mathbf R)=\mathbf W^{1/2}\mathbf B(\mathbf R)
\)
and
\(
\mathbf Q_v(\mathbf R)=\mathbf C^{\herm}(\mathbf R)\mathbf C(\mathbf R)
\).
The current regularization also improves
numerical conditioning by making
$\mathbf A^{\herm}\mathbf A+\rho_i\mathbf I$ positive definite.
Eliminating the current from the joint design gives the reduced objective
\begin{equation}
\begin{aligned}
 F(\mathbf R)=\min_{\mathbf i}\quad&
 \norm{\mathbf A(\mathbf R)\mathbf i-\widetilde{\mathbf g}}_2^2
 {}+\rho_i\norm{\mathbf i}_2^2\\
 \mathrm{s.t.}\quad&
 \mathbf i^{\herm}\mathbf Q_v(\mathbf R)\mathbf i\le V_{\max}^2 .
\end{aligned}
\label{eq:fixed_current}
\end{equation}
This is a convex quadratically constrained quadratic program.  Its KKT
solution is
\begin{equation}
 \mathbf i(\mu)=\left(\mathbf A^{\herm}\mathbf A+\rho_i\mathbf I
 +\mu\mathbf Q_v\right)^{-1}\mathbf A^{\herm}\widetilde{\mathbf g},
\label{eq:current}
\end{equation}
where $\mu\ge0$ is the multiplier of the source-voltage constraint.  If the
unconstrained current is feasible, then $\mu=0$.  Otherwise, scalar bisection
selects $\mu>0$ such that
$\norm{\mathbf C(\mathbf R)\mathbf i(\mu)}_2=V_{\max}$.  The implementation
whitens the problem by $\mathbf C(\mathbf R)$ and diagonalizes the resulting
Hermitian matrix once per geometry, leaving only scalar and vector operations
inside the bisection.

\subsection{EM-Aware Reduced Position Gradient}

Let $p_{n,1}=x_n$, $p_{n,2}=y_n$, and let
\(
\mathbf e=\mathbf W^{1/2}(\mathbf B\mathbf i-\mathbf g)
\)
be the weighted residual evaluated at the exact current from
\eqref{eq:fixed_current}.  The response derivatives for angular sample
$\ell$ are
\begin{equation}
 \frac{\partial B_{\ell,n}}{\partial x_n}
 =-\junit2\pi u_\ell B_{\ell,n},\qquad
 \frac{\partial B_{\ell,n}}{\partial y_n}
 =-\junit2\pi v_\ell B_{\ell,n}.
\label{eq:response_derivatives}
\end{equation}
For $m\ne n$, the position-dependent mutual-impedance entries satisfy
\begin{equation}
 \frac{\partial C_{n,m}}{\partial p_{n,k}}
 =z'(d_{n,m})\frac{p_{n,k}-p_{m,k}}{d_{n,m}},
 \qquad k\in\{1,2\},
\label{eq:coupling_derivative}
\end{equation}
with the same derivative in the reciprocal entry $C_{m,n}$.  The diagonal,
loss, and source-impedance terms are position independent.

Applying the envelope theorem to~\eqref{eq:fixed_current} yields
\begin{equation}
 \frac{\partial F}{\partial p_{n,k}}=
 2\Re\!\left\{\mathbf e^{\herm}\mathbf W^{1/2}
 \frac{\partial\mathbf B}{\partial p_{n,k}}\mathbf i\right\}
 +2\mu\Re\!\left\{\mathbf v_s^{\herm}
 \frac{\partial\mathbf C}{\partial p_{n,k}}\mathbf i\right\}.
\label{eq:gradient}
\end{equation}
The first term captures the position-dependent radiation response.  The
second captures the change in the source-voltage feasible set caused by
mutual coupling, and it vanishes when the voltage constraint is inactive.
Derivatives of the optimized current do not appear because the current
stationarity conditions are already satisfied.

\subsection{Projected Adam}

Let $\nabla F\in\R^{N_a\times2}$ collect~\eqref{eq:gradient}.  At iteration
$t$, Adam forms the elementwise moments~\cite{kingma2015adam}
\begin{equation}
\begin{aligned}
 \mathbf m_t&=\beta_1\mathbf m_{t-1}+(1-\beta_1)\nabla F,\\
 \mathbf s_t&=\beta_2\mathbf s_{t-1}
 +(1-\beta_2)(\nabla F\odot\nabla F),
\end{aligned}
\label{eq:adam_moments}
\end{equation}
and their bias-corrected versions
$\widehat{\mathbf m}_t$ and $\widehat{\mathbf s}_t$.  The tentative update is
\begin{equation}
 \widetilde{\mathbf R}_t=\mathbf R_{t-1}
 -\alpha\frac{\widehat{\mathbf m}_t}
 {\sqrt{\widehat{\mathbf s}_t}+\epsilon},
\label{eq:adam_position}
\end{equation}
where the division and square root are elementwise.  Before feasibility
repair, the largest per-port trial displacement is capped at
$\Delta_{\rm step}$.

Each tentative geometry then undergoes three cyclic operations.  Aperture
projection clips every coordinate to $[0,L]$.  Travel projection maps each
$\mathbf r_n$ into the radius-$D$ disk centered at
$\mathbf r_n^{(0)}$.  Spacing repair moves every violating pair apart
symmetrically until its separation reaches $d_{\min}$.
The operations repeat until all constraints pass.  Their composite map is
denoted by $\mathcal P_{\rm ap,tr,sp}$.  Since the spacing set is nonconvex,
this map is a deterministic feasibility repair rather than a globally nearest
projection.  Every feasible trial geometry rebuilds
$\mathbf Z_{\rm em}(\mathbf R)$ and $\mathbf C(\mathbf R)$, resolves
\eqref{eq:fixed_current}, and refreshes~\eqref{eq:gradient}.  Projected Adam
need not decrease $F$ at every iteration, so the algorithm returns the
feasible geometry with the smallest visited reduced objective.

\section{Numerical Results}\label{sec:results}
\begin{figure*}[!t]
\centering
\includegraphics[width=0.81\textwidth]{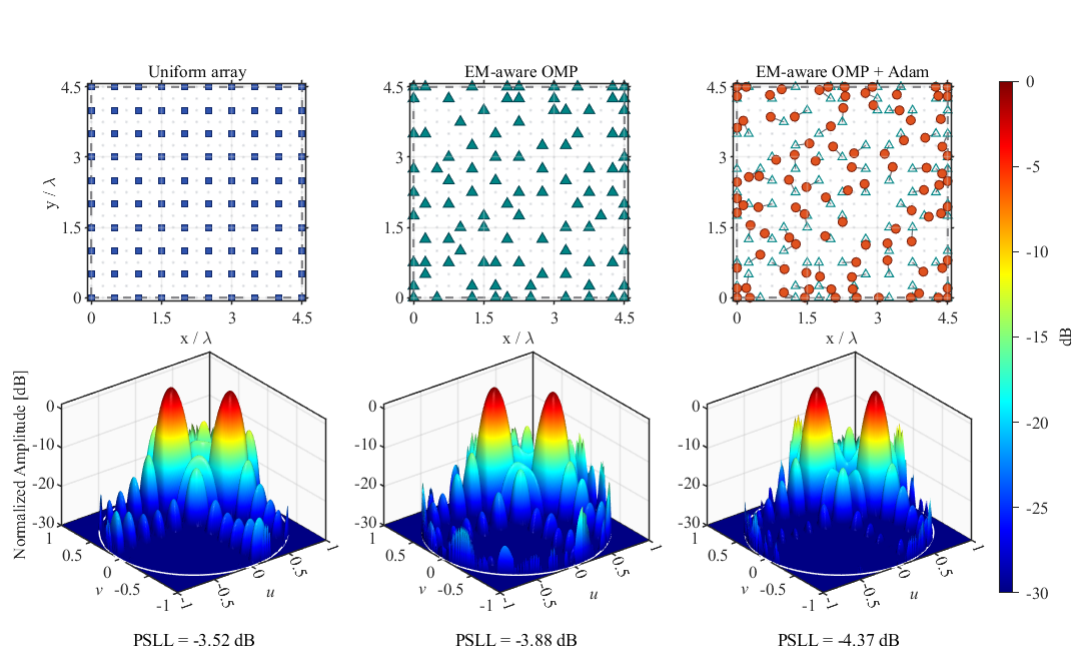}
\caption{Representative port layouts and three-dimensional beam patterns.
Columns correspond to the uniform array, EM-aware OMP, and the proposed
method.  Top row, active-port locations, with open triangles and filled
circles in the right panel denoting the initial and refined positions and
gray arrows their displacement.  Bottom row, independently peak-normalized
patterns over the visible direction-cosine disk $u^2+v^2\leq1$, clipped to
$[-30,0]$~dB.  All currents are re-solved under the true coupled network.}
\label{fig:heatmaps}
\vspace{-4mm}
\end{figure*}
\subsection{Protocol and Metrics}

Table~\ref{tab:parameters} lists the nominal configuration.  The desired
amplitude $\mathbf g_{\mathrm{am}}$ contains
two symmetric beams on a common elevation cut.  At direction $(\theta,\phi)$,
beam $b$ has center $(\theta_b,\phi_b)$, elevation/azimuth half widths
$(\Delta\theta_b,\Delta\phi_b)$, peak amplitude $a_b$, and normalized radius
\begin{equation}
 \rho_b=\max\!\left\{\frac{|\theta-\theta_b|}{\Delta\theta_b},
 \frac{|\operatorname{wrap}(\phi-\phi_b)|}{\Delta\phi_b}\right\},
\end{equation}
where $\operatorname{wrap}(\cdot)$ returns the principal angular difference.
For flat-top ratio $\kappa_b$, the amplitude equals $a_b$ when
$\rho_b\le\kappa_b$, decreases linearly to zero over
$\kappa_b<\rho_b\le1$, and is zero otherwise.  The nominal beams have
centers $(45^\circ,20^\circ)$ and $(45^\circ,70^\circ)$, common half widths
$(8^\circ,8^\circ)$, flat-top ratio 0.45, and unit amplitude.

\begin{table}[!t]
\centering
\caption{Nominal Simulation Parameters}
\label{tab:parameters}
\footnotesize
\renewcommand{\arraystretch}{1.05}
\begin{tabular}{@{}
>{\raggedright\arraybackslash}p{0.38\columnwidth}
>{\centering\arraybackslash}p{0.22\columnwidth}
>{\raggedright\arraybackslash}p{0.30\columnwidth}@{}}
\toprule
Parameter & Symbol & Value \\
\midrule
Physical antennas & $N_a$ & $100$ \\
OMP candidates & $N_p$ & $361$ \\
Aperture & $L\times L$ & $4.5\lambda\times4.5\lambda$ \\
Candidate spacing & $d_{\rm cand}$ & $0.25\lambda$ \\
Minimum spacing & $d_{\min}$ & $0.2\lambda$ \\
Travel / voltage budget & $D,V_{\max}$ & $0.3\lambda,35$ \\
Grid side lengths & $N_{\rm d},N_{\rm e}$ & $61,121$ \\
Weight parameters & $w_{\rm side},\gamma$ & $12,1.2$ \\
Adam rate / step cap & $\alpha,\Delta_{\rm step}$ & $0.016\lambda,0.06\lambda$ \\
Adam moments / iterations & $\beta_1,\beta_2,T_{\max}$ & $0.9,0.999,100$ \\
Noise / EM loading & $\sigma_n^2,\rho_z$ & $10^{-3},0.05~\Omega$ \\
\bottomrule
\end{tabular}
\end{table}

The evaluation grid has $N_{\rm e}$ samples per axis and
$M_{\rm e}=N_{\rm e}^2$ directions.  Let
$\mathbf y=[y_1,\ldots,y_{M_{\rm e}}]^\trans$ denote the synthesized field
and define the mainlobe and sidelobe sample sets as
\begin{equation}
 \mathcal M=\{m\mid[\mathbf g_{\rm am}]_m\ge0.5\},\qquad
 \mathcal S=\{m\mid[\mathbf g_{\rm am}]_m\le0.05\}.
\end{equation}
Following the average-mainlobe and peak-sidelobe criteria, the two reported
metrics are the average mainlobe signal-to-noise ratio (SNR) and the peak
sidelobe level (PSLL):
\begin{align}
 \mathrm{SNR}_{\rm avg}
 &=10\log_{10}\!\left(
 \frac{1}{|\mathcal M|\sigma_n^2}
 \sum_{m\in\mathcal M}|y_m|^2\right),\label{eq:metric-snr}\\
 \mathrm{PSLL}
 &=20\log_{10}\!\left(
 \frac{\max_{m\in\mathcal S}|y_m|}
 {\max_m|y_m|}\right).\label{eq:metric-psll}
\end{align}
We compare
a fixed $10\times10$ uniform array, grid-based EM-aware OMP, and the proposed
two-stage method, which runs EM-aware OMP followed by projected Adam.  The
first two geometries are static.  All three
methods re-solve the voltage-constrained current subproblem and evaluate
their fields with the true coupled network.

At $D=0.3\lambda$, 24 held-out targets give median average-mainlobe SNR and
PSLL improvements of 1.25 and 0.49~dB over EM-aware OMP, with both gains
positive in all cases.

\subsection{Beam-Pattern Comparison}

The top row of Fig.~\ref{fig:heatmaps} compares the uniform lattice, 100 ports
selected by EM-aware OMP, and their projected-Adam refinement.  The mean,
90th-percentile, and maximum movements are $0.19\lambda$, $0.30\lambda$, and
$0.30\lambda$, while the final minimum spacing is $0.20\lambda$ and the
source-voltage norm reaches 35.  The nonuniform displacements rule out a rigid
aperture translation.

The bottom row confirms the requested dual beams.  The uniform, EM-aware OMP,
and projected-Adam PSLL values are $-3.52$, $-3.88$, and $-4.37$~dB.
Relative to EM-aware OMP, projected Adam improves average-mainlobe SNR and
PSLL by 1.33 and 0.49~dB; the gains over the uniform array are 2.44 and
0.85~dB.  The plots are independently peak-normalized to emphasize shape,
whereas SNR uses the unnormalized coupled-network fields.  The gains therefore
reflect reduced grid mismatch rather than array-factor scaling.

\begin{figure}[!t]
\centering
\includegraphics[width=0.99\columnwidth]{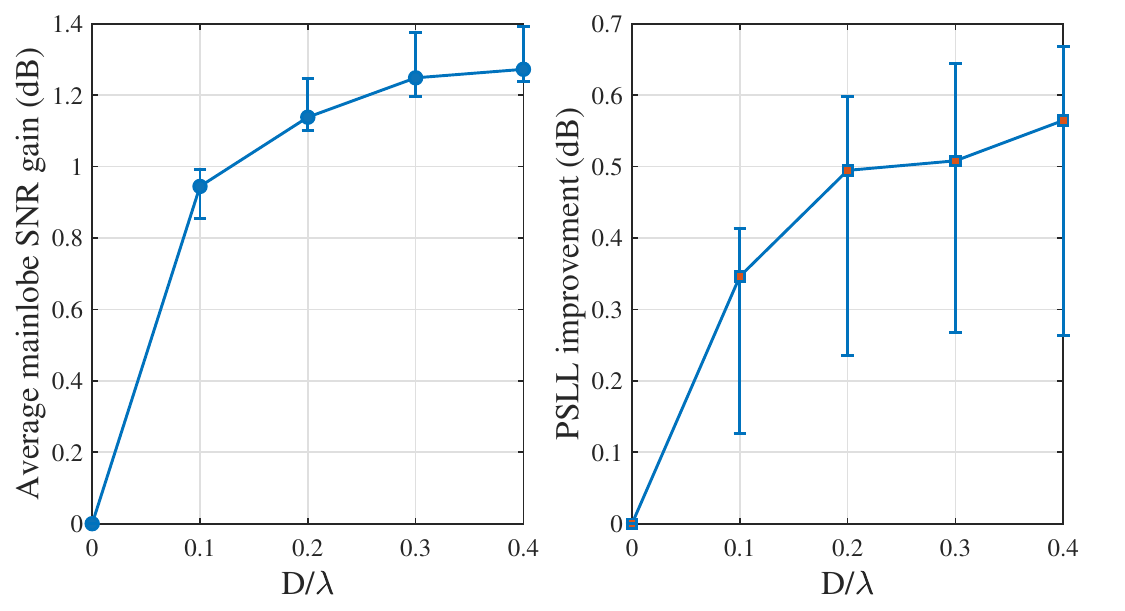}
\caption{Median average-mainlobe SNR gain and PSLL improvement relative to
$D=0$.  Error bars show interquartile ranges over 20 independently perturbed
movement-budget targets, each initialized by EM-aware OMP.}
\label{fig:budget}
\vspace{-4mm}
\end{figure}
\subsection{Movement-Budget Tradeoff}

Fig.~\ref{fig:budget} summarizes 20 independently perturbed dual-beam targets,
each with its own EM-aware OMP initialization.  From $0.10\lambda$ to
$0.30\lambda$, median average-mainlobe SNR gain rises from 0.94 to 1.25~dB and
median PSLL improvement from 0.35 to 0.51~dB, with positive lower quartiles at
every nonzero budget.  Raising the budget to $0.40\lambda$ adds only 0.02~dB
SNR and 0.05~dB PSLL while median actual movement rises from $0.20\lambda$ to
$0.22\lambda$, confirming saturation beyond $0.30\lambda$.

\section{Conclusion}\label{sec:conclusion}
This paper combines phase retrieval, EM-aware OMP, and movement-constrained projected Adam for FAA beamforming with position-dependent impedance. Across 24 held-out validation targets at $D=0.3\lambda$, projected Adam improves median average-mainlobe SNR and PSLL by 1.25 and 0.49~dB, with both gains positive in every case. Gains persist for all nonzero budgets and saturate beyond $0.3\lambda$. Future work will incorporate measured/full-wave data, model uncertainty, and hardware motion dynamics.
\balance
\bibliographystyle{IEEEtran}
\bibliography{references}

@IEEEtranBSTCTL{IEEEexample:BSTcontrol,
		CTLdash_repeated_names = "no",
  CTLuse_forced_etal = {yes},
  CTLmax_names_forced_etal = {6},
  CTLnames_show_etal = {1}
}

@Article{wong2020limits,
  author   = {Wong, Kai-Kit and others},
  journal  = {IEEE Commun. Lett.},
  title    = {Performance Limits of Fluid Antenna Systems},
  year     = {2020},
  month    = {Nov.},
  number   = {11},
  pages    = {2469--2472},
  volume   = {24},
  doi      = {10.1109/LCOMM.2020.3006554},
  fjournal = {IEEE Communications Letters},
}

@Article{wong2021fas,
  author   = {Wong, Kai-Kit and others},
  journal  = {IEEE Trans. Wireless Commun.},
  title    = {Fluid Antenna Systems},
  year     = {2021},
  month    = {Mar.},
  number   = {3},
  pages    = {1950--1962},
  volume   = {20},
  doi      = {10.1109/TWC.2020.3037595},
  fjournal = {IEEE Transactions on Wireless Communications},
}

@ARTICLE{CSI1,
	author  = {Zhentian Zhang and others},
	title   = {Joint Activity Detection and Channel Estimation for Fluid Antenna System Exploiting Geographical and Angular Information},
	journal = {IEEE J. Sel. Topics Signal Process.},
	year    = {2026},
	volume  = {20},
	number  = {3},
	pages   = {354--370},
	doi     = {10.1109/JSTSP.2026.3673148}
}

@ARTICLE{CSI2,
	author  = {Y. Wu and others},
	title   = {Learned-Approximate Message Passing Under {Karhunen--Lo\`eve} Modeling for Fluid Antenna Systems},
	journal = {IEEE Wireless Commun. Lett.},
	year    = {2026},
	volume  = {15},
	pages   = {2719--2723},
	doi     = {10.1109/LWC.2026.3682568}
}

@article{CSI3,
	author  = {Z. Zhang and others},
	title   = {Beyond Covariance: Generative Spatial Correlation Modeling and Channel Interpolation for Fluid Antenna Systems},
	journal = {arXiv preprint},
	year    = {2026},
	url     = {https://arxiv.org/abs/2604.16639}
}

@inproceedings{CSI4,
	author    = {Z. Zhang and others},
	title     = {Low-Complexity {CSI} Acquisition Exploiting Geographical Diversity in Fluid Antenna System},
	booktitle = {Proc. IEEE Global Commun. Conf. Workshops (GC Wkshps)},
	year      = {2025},
	pages     = {308--313}
}

@Article{CSI5,
	author  = {Z. Zhang and others},
	title   = {Geometry-Structured Channel Reconstruction for Conventional and Fluid Antenna Systems: Bayesian Inference and Fundamental Limits},
	journal = {arXiv preprint},
	year    = {2026},
	url     = {https://arxiv.org/abs/2606.04001}
}

@Article{zhang2026finite,
  author   = {Zhang, Zhentian and others},
  journal  = {IEEE Wireless Commun. Lett.},
  title    = {Finite-Aperture Fluid Antenna Array Design: Analysis and Algorithm},
  year     = {2026},
  pages    = {3199--3203},
  volume   = {15},
  doi      = {10.1109/LWC.2026.3685911},
  fjournal = {IEEE Wireless Communications Letters},
}

@Article{zhang2026fbl,
  author   = {Zhang, Zhentian and Wong, Kai-Kit and Morales-Jimenez, David and Jiang, Hao and Xu, Hao and Masouros, Christos and Zhang, Zaichen},
  journal  = {IEEE Trans. Wireless Commun.},
  title    = {Finite-blocklength Fluid Antenna Systems},
  year     = {2026},
  note     = {early access},
  doi      = {10.1109/TWC.2026.3723456},
  fjournal = {IEEE Transactions on Wireless Communications}
}

@INPROCEEDINGS{11555734,
  author={Xu, Jingyuan and others},
  booktitle={2026 IEEE Wireless Communications and Networking Conference Workshops (WCNCW)},
  title={Fluid Antenna-Enhanced Flexible Beamforming},
  year={2026},
  volume={},
  number={},
  pages={1-6},
  doi={10.1109/WCNCW67598.2026.11555734}}

@Article{yang2022sparse,
  author   = {Yang, Songjie and others},
  journal  = {IEEE Antennas Wirel. Propag. Lett.},
  title    = {Low-Complexity Sparse Array Synthesis Based on Off-Grid Compressive Sensing},
  year     = {2022},
  month    = {Dec.},
  number   = {12},
  pages    = {2322--2326},
  volume   = {21},
  doi      = {10.1109/LAWP.2022.3192308},
  fjournal = {IEEE Antennas and Wireless Propagation Letters},
}

@MISC{zhang2026emaware,
  author={Zhang, Zhentian and others},
title={Electromagnetic-Aware Fluid Antenna Array},
year={2026},
eprint={2607.21375},
archivePrefix={arXiv},
url={https://arxiv.org/abs/arXiv:2607.21375}
}

@Article{tropp2007omp,
  author   = {Tropp, Joel A. and Gilbert, Anna C.},
  journal  = {IEEE Trans. Inf. Theory},
  title    = {Signal Recovery From Random Measurements Via Orthogonal Matching Pursuit},
  year     = {2007},
  month    = {Dec.},
  number   = {12},
  pages    = {4655--4666},
  volume   = {53},
  doi      = {10.1109/TIT.2007.909108},
  fjournal = {IEEE Transactions on Information Theory},
}

@ARTICLE{gerchberg1972,
  author={Gerchberg, R. W. and Saxton, W. O.},
  journal={Optik},
  title={A Practical Algorithm for the Determination of Phase From Image and Diffraction Plane Pictures},
  year={1972},
  volume={35},
  number={2},
  pages={237--246}
}

@Article{echeveste2016,
  author   = {Echeveste, Jos{\'e} Ignacio and others},
  journal  = {IEEE Trans. Antennas Propag.},
  title    = {Near-Optimal Shaped-Beam Synthesis of Real and Coupled Antenna Arrays via 3-D-FEM and Phase Retrieval},
  year     = {2016},
  month    = {Jun.},
  number   = {6},
  pages    = {2189--2196},
  volume   = {64},
  doi      = {10.1109/TAP.2016.2546299},
  fjournal = {IEEE Transactions on Antennas and Propagation},
}

@Article{gupta1983mutual,
  author   = {Gupta, Inder J. and Ksienski, Arthur A.},
  journal  = {IEEE Trans. Antennas Propag.},
  title    = {Effect of Mutual Coupling on the Performance of Adaptive Arrays},
  year     = {1983},
  month    = {Sep.},
  number   = {5},
  pages    = {785--791},
  volume   = {31},
  doi      = {10.1109/TAP.1983.1143128},
  fjournal = {IEEE Transactions on Antennas and Propagation},
}

@BOOK{balanis2016,
  author={Balanis, Constantine A.},
  title={Antenna Theory: Analysis and Design},
  edition={4th},
  publisher={Wiley},
  address={Hoboken, NJ, USA},
  year={2016}
}

@Article{chen2026nearfield,
  author   = {Chen, Sen and others},
  journal  = {IEEE Commun. Lett.},
  title    = {Near-Field Beamforming and Port Selection for Fluid Antenna Systems},
  year     = {2026},
  pages    = {2248--2252},
  volume   = {30},
  fjournal = {IEEE Communications Letters},
}

@Article{zhang2011robust,
  author   = {Zhang, Tongtong and Ser, Wee},
  journal  = {IEEE Trans. Antennas Propag.},
  title    = {Robust Beampattern Synthesis for Antenna Arrays With Mutual Coupling Effect},
  year     = {2011},
  month    = {Aug.},
  number   = {8},
  pages    = {2889--2895},
  volume   = {59},
  doi      = {10.1109/TAP.2011.2152329},
  fjournal = {IEEE Transactions on Antennas and Propagation},
}

@ARTICLE{liang2026peaksidelobe,
  author={H. Liang and others},
  title={Peak Sidelobe Suppression in Planar Fluid Antenna Array},
  journal={arXiv preprint},
  year={2026},
  url={https://arxiv.org/abs/2606.31149}
}

@INPROCEEDINGS{kingma2015adam,
  author={Kingma, Diederik P. and Ba, Jimmy},
  title={Adam: A Method for Stochastic Optimization},
  booktitle={International Conference on Learning Representations (ICLR)},
  year={2015}
}

@ARTICLE{hj3,
  author={Hong, H. and others},
  journal={IEEE Open Journal of the Communications Society},
  title={Fluid Antenna Multiple Access for {6G}: A Holistic Review},
  year={2026},
  volume={7},
  pages={2607--2633}
}

@Article{nb1,
  author   = {Wang, Z. and others},
  journal  = {IEEE Trans. Veh. Technol.},
  title    = {Optimal Bilinear Equalizer Beamforming Design for Cell-Free Massive {MIMO} Networks With Arbitrary Channel Estimators},
  year     = {2025},
  month    = {Apr.},
  number   = {4},
  pages    = {6862--6867},
  volume   = {74},
  fjournal = {IEEE Transactions on Vehicular Technology},
}

@ARTICLE{nb2,
  author={Girnyk, M. and others},
  journal={Ericsson Technology Review},
  title={Broad Beamforming Technology in {5G} Massive {MIMO}},
  year={2023},
  volume={2023},
  number={10},
  pages={2--6},
  month={Oct.}
}

@Article{wt2,
  author   = {Wu, T. and others},
  journal  = {IEEE J. Sel. Topics Signal Process.},
  title    = {Scalable Fluid Antenna Systems: A New Paradigm for Array Signal Processing},
  year     = {2026},
  month    = {Apr.},
  number   = {3},
  pages    = {389--406},
  volume   = {20},
  fjournal = {IEEE Journal of Selected Topics in Signal Processing},
}

@Article{wt5,
  author   = {Wu, Tuo and Zhi, Kangda and Yao, Junteng and Lai, Xiazhi and Zheng, Jianchao and Niu, Hong and Elkashlan, Maged and Wong, Kai-Kit and Chae, Chan-Byoung and Ding, Zhiguo and Karagiannidis, George K. and Debbah, M{\'e}rouane and Yuen, Chau},
  journal  = {IEEE Wirel. Commun.},
  title    = {Fluid Antenna Systems Enabling {6G}: Principles, Applications, and Research Directions},
  year     = {2026},
  number   = {4},
  pages    = {100--108},
  volume   = {33},
  fjournal = {IEEE Wireless Communications},
}

@MISC{zhang2026planar,
  author={Zhang, Zhentian and others},
  title={Finite-Aperture Planar Fluid Antenna Array},
  year={2026},
  month={May},
  eprint={2605.22040},
  archivePrefix={arXiv},
 url = {https://arxiv.org/abs/arXiv:2605.22040}
}

@Article{zzt1,
  author   = {Zhang, Z. and others},
  journal  = {IEEE J. Sel. Areas Commun.},
  title    = {On Fundamental Limits for Fluid Antenna-Assisted Integrated Sensing and Communications for Unsourced Random Access},
  year     = {2026},
  pages    = {136--149},
  volume   = {44},
  fjournal = {IEEE Journal on Selected Areas in Communications},
}

@Article{zzt_wcl1,
  author   = {Zhang, Zhentian and others},
  journal  = {IEEE Wireless Commun. Lett.},
  title    = {Jointly Correlated Dual-Side Fluid Antenna System},
  year     = {2026},
  pages    = {4370-4374},
  volume   = {15},
  doi      = {10.1109/LWC.2026.3710907},
  fjournal = {IEEE Wireless Communications Letters},
}

@Article{zzt_wcl2,
  author   = {Zhang, Zhentian and others},
  journal  = {IEEE Wireless Commun. Lett.},
  title    = {Cramér–Rao Bounds for Activity Detection in Conventional and Fluid Antenna Systems},
  year     = {2026},
  pages    = {3059-3063},
  volume   = {15},
  doi      = {10.1109/LWC.2026.3691385},
  fjournal = {IEEE Wireless Communications Letters},
}

@Article{zzt_wcl3,
  author   = {Zhang, Zhentian and others},
  journal  = {IEEE Wireless Commun. Lett.},
  title    = {Finite-Blocklength Fluid Antenna Systems With Spatial Block-Correlation Channel Model},
  year     = {2026},
  pages    = {1911-1915},
  volume   = {15},
  doi      = {10.1109/LWC.2026.3666168},
  fjournal = {IEEE Wireless Communications Letters},
}

@Article{zzt_wcl4,
  author   = {Zhang, Zhentian and others},
  journal  = {IEEE Wireless Commun. Lett.},
  title    = {On Fundamental Limits of Slow-Fluid Antenna Multiple Access for Unsourced Random Access},
  year     = {2025},
  number   = {11},
  pages    = {3455-3459},
  volume   = {14},
  doi      = {10.1109/LWC.2025.3594112},
  fjournal = {IEEE Wireless Communications Letters},
}
\end{document}